\documentclass[12pt]{article}
\pdfoutput=1

\author{Andrew S. Goetz}
\title{Tully-Fisher Scalings and Boundary Conditions for Wave Dark Matter}

\usepackage{hyperref}

\usepackage{amsmath,amssymb,amsthm}
\usepackage[backend=bibtex8,hyperref=auto,style=numeric]{biblatex}
\usepackage{cleveref}
\usepackage{enumitem}
\usepackage[letterpaper,left=1.2in,right=1.2in,top=1.2in,bottom=1.2in,marginparwidth=1.5in]{geometry}
\usepackage{mathabx}
\usepackage{mathtools}
\usepackage{microtype}
\usepackage{pgfplots}
  \pgfplotsset{compat=newest}
\usepackage{siunitx}
\usepackage{titling}
\theoremstyle{definition}
\newtheorem{definition}{Definition}
\AfterEndEnvironment{definition}{\noindent\ignorespaces}
\newtheorem{question}{Question}
\AfterEndEnvironment{question}{\noindent\ignorespaces}

\DeclarePairedDelimiter\abs{\lvert}{\rvert} % absolute value
\newcommand\RDM{R_{\text{DM}}}
\newcommand\tensor{\otimes}

\newcommand\Vtil{\tilde{V}}

\DeclareMathOperator\Ai{Ai}
\DeclareMathOperator\Bi{Bi}

\DeclareSIUnit\solarmass{\ensuremath{M_\Sun}}
\DeclareSIUnit\ly{ly}

\newread\myread

\begin{document}

\maketitle
\begin{abstract}
We investigate a theory of dark matter called wave dark matter, also known as scalar field dark matter (SFDM) and boson star dark matter or Bose-Einstein condensate (BEC) dark matter (also see axion dark matter), and its relation to the Tully-Fisher relation. We exhibit two boundary conditions that give rise to Tully-Fisher-like relations for spherically symmetric static wave dark matter halos: (BC1) Fixing a length scale at the outer edge of wave dark matter halos gives rise to a Tully-Fisher-like relation of the form $M/v^4=\text{constant}$. (BC2) Fixing the density of dark matter at the outer edge of wave dark matter halos gives rise to a Tully-Fisher-like relation of the form $M/v^{3.4}=\text{const}$. These results extend the work of a previous paper \cite{bray14}.
\end{abstract}

\section{Introduction} \label{sec:intro}
The baryonic Tully-Fisher relation (BTFR) is an empirical scaling relation for disk galaxies involving the baryonic (i.e., non-dark) mass $M_b$ and the rotational velocity $v$ of stars. It is observed that
\begin{equation} \label{eq:BTFR1}
\frac{M_b}{v^x} = \text{const.} 
\end{equation}
where the exponent $x$ (the slope of the relation in log-log space) is usually found to lie between $3$ and $4$. There is an extensive literature which uses astronomical data to calibrate the relation and estimate $x$. See, for example, \cite{mcgaugh00,bell01,verheijen01,mcgaugh05,geha06,gurovich10,mcgaugh10,torresflores11,mcgaugh12}.

In the relation \eqref{eq:BTFR1}, the rotational velocity $v$ is determined by the total gravitating mass (baryonic and dark) while $M_b$ accounts for the baryonic matter only. Thus, the BTFR provides a link between the baryonic and dark matter and may have something to say about the viability of various theories of dark matter.

In this paper we discuss the BTFR in the context of a theory of wave dark matter. Wave dark matter \cite{bray13b,bray12,bray13a,bray14,parry12b,parrythesis} has been investigated under other names such as scalar field dark matter (SFDM) \cite{guzman00,guzman04,magana12,suarez14,sahni00} and boson star dark matter or Bose-Einstein condensate (BEC)  dark matter \cite{seidel90,sin94,ji94,lai07,matos07,sharma08,urenalopez10}. The difference in names comes from a difference in motivations, but the underlying equation is the Klein-Gordon wave equation \eqref{eq:ekg2} for a scalar field.

In spherical symmetry, there are simple ``static states'' of wave dark matter (see \cref{fig:0,fig:1,fig:10,fig:100} at the end of the paper). Sin proposed in \cite{sin94} that each galactic dark matter halo corresponds (in a first approximation) to one of these static states. If we take this suggestion as a starting point, a natural concomitant idea is that the baryonic Tully-Fisher relation might arise because a corresponding ``Tully-Fisher-like'' relation holds for the wave dark matter halos. This Tully-Fisher-like relation would hold because of the nature of wave dark matter. The fundamental question we ask in this paper is whether the equations of wave dark matter have the potential to give rise to such a Tully-Fisher-like relation (see \cref{q:tfscalingsimprecise} on \cpageref{q:tfscalingsimprecise}). We answer this question in the affirmative and exhibit two boundary conditions, which could be physical conditions imposed at the edge of each wave dark matter halo, that give rise to just such a Tully-Fisher-like relation. The boundary conditions are as follows:
\begin{description}[style=multiline,leftmargin=1.5cm]
\item[BC1:] Fixing a length scale at the outer edge of halos implies a Tully-Fisher-like relation with slope $4$. This is a more general version of a result from a previous paper \cite{bray14}.
\item[BC2:] Fixing the dark matter density at the outer edge of halos implies a Tully-Fisher-like relation with slope $3.4$.
\end{description}
The remainder of this paper runs as follows: In \cref{sec:ssss} we introduce the spherically symmetric static states and their scaling properties. In \cref{sec:numerical} we exhibit \textbf{BC1} and \textbf{BC2} numerically, and in \cref{sec:theoretical} we derive \textbf{BC1} and \textbf{BC2} theoretically. In \cref{sec:other_scalings} we describe other conditions one might impose which do not lead to Tully-Fisher-like relations. Finally in \cref{sec:conclusion} we briefly discuss the connection between a Tully-Fisher-like relation for wave dark matter halos and the baryonic Tully-Fisher relation.

\section{The Spherically Symmetric Static States} \label{sec:ssss}
The equations underlying the theory of wave dark matter are the Einstein-Klein-Gordon (EKG) equations:
\begin{subequations} \label{eq:ekg}
\begin{align}
G &= 8\pi\left( \frac{df\tensor d\bar{f}+d\bar{f}\tensor df}{\Upsilon^2} - \left(\frac{\abs{df}^2}{\Upsilon^2}+\abs{f}^2\right)g\right) \label{eq:ekg1} \\
\Box_g f &= \Upsilon^2 f. \label{eq:ekg2}
\end{align}
\end{subequations}
(Throughout, we use geometrized units in which the universal gravitation constant and the speed of light are unity: $G=c=1$.) Here $f$ is a complex scalar field representing the dark matter and $\Upsilon$ is a fundamental constant of nature. The relationship between $\Upsilon$ and the wave dark matter particle mass $m$ is
\begin{equation}
m = \frac{\hbar\Upsilon}{c} = \SI{2.09e-23}{eV}\left(\frac{\Upsilon}{\SI{1}{\ly^{-1}}}\right).
\end{equation}
In this paper for definiteness we use $\Upsilon=\SI{10}{\ly^{-1}}$, which is consistent with constraints from other studies \cite{hu00,bray13a,suarez14}.

Because we are interested in studying galactic dark matter halos, most of which are thought to be approximately spherically symmetric, we study the EKG equations in spherical symmetry. The spacetime metric we will use is
\begin{equation} \label{eq:metric}
g = -e^{2V(t,r)}\,dt^2 + \Phi(t,r)^{-1}\,dr^2 + r^2(d\theta^2 + \sin^2\phi\,d\phi^2)
\end{equation}
where $\Phi(t,r)=1-\frac{2M(t,r)}{r}$. In the non-relativistic limit, the functions $V(t,r)$ and $M(t,r)$ have natural interpretations as the gravitational potential at time $t$ and radius $r$ and the total mass contained inside a ball of radius $r$ at time $t$. The simplest solutions to \cref{eq:ekg} in the metric \eqref{eq:metric} are the \emph{spherically symmetric static states}, which are of the form
\begin{equation} \label{eq:staticansatz}
f(t,r)=F(r)e^{i\omega t}.
\end{equation}
Substituting the ansatz \eqref{eq:staticansatz} into \cref{eq:ekg} and using \cref{eq:metric}, we obtain (see \cite{parrythesis}) the relatively simple system of ODEs
\begin{subequations} \label{eq:ekgode}
\begin{gather}
M_r = 4\pi r^2\cdot\frac{1}{\Upsilon^2}\left[ \left(\Upsilon^2+\omega^2e^{-2V}\right)F^2 + \Phi F_r^2\right]\label{eq:ekgodeM} \\
\Phi V_r = \frac{M}{r^2} - 4\pi r\cdot\frac{1}{\Upsilon^2}\left[ \left(\Upsilon^2-\omega^2 e^{-2V}\right)F^2 - \Phi F_r^2 \right] \label{eq:ekgodeV} \\
F_{rr} + \frac{2}{r} F_r + V_rF_r + \frac{\Phi_r}{\Phi}F_r = \Phi^{-1}\left(\Upsilon^2-\omega^2e^{-2V}\right)F \label{eq:ekgodeF}
\end{gather}
\end{subequations}
which can be integrated on a computer. Here the functions $M(r),V(r),F(r)$ are functions of $r$ only, and the only dependence on $t$ appears in \cref{eq:staticansatz}.

In the non-relativistic limit the solutions of \eqref{eq:ekgode} are indistinguishable from the solutions of the even simpler system of ODEs
\begin{subequations} \label{eq:psode}
\begin{align}
M_r &= 4\pi r^2\cdot 2F^2 \label{eq:psodeM} \\
V_r &= \frac{M}{r^2} \label{eq:psodeV} \\
\frac{1}{2\Upsilon}\left(F_{rr} + \frac{2}{r} F_r\right) &= (\Upsilon-\omega+\Upsilon V)F. \label{eq:psodeF}
\end{align}
\end{subequations}
This system is the Poisson-Schr\"odinger (PS) system written in spherical symmetry. Here we have a special case of the well-known fact that the PS system is the non-relativistic limit of the EKG system---see \cite{giulini12}. From here on in this paper we work exclusively with the simpler PS system \eqref{eq:psode} because we are interested in wave dark matter in galaxies which are, considered as an aggregate, non-relativistic systems.

Finite-mass solutions to \cref{eq:psode} come in the form of ground states and excited states. It was proposed in \cite{sin94} that each galactic dark matter halo corresponds (in a first approximation) to one of these states. In \cref{fig:0,fig:1,fig:10,fig:100} we display some representative graphs of solutions so that the reader can get a sense of what they look like, and refer the reader to a previous paper \cite{bray14} for more examples. We assign each solution a non-negative integer $n$ which counts the number of zeros (nodes) of $F$. A ground state has $n=0$; excited states have $n\geq 1$. Any solution can be given as a quadruple $(\omega;M,V,F)$ consisting of the constant $\omega$ followed by the three functions $M(r)$, $V(r)$, and $F(r)$. It is easily verified that if $(\omega;M,V,F)$ is a solution, then so is $(\omega+\Upsilon\Vtil;M,V+\Vtil,F)$. We use this freedom to shift every potential function $V(r)$ so that it approaches zero at infinity.

For each static state there is a special value of $r$ which we refer to as $\RDM$ and which we think of as the edge of the dark matter halo. Mathematically, $\RDM$ is the particular value of $r$ at which the function $F(r)$ switches from oscillatory to exponentially decreasing behavior. Looking at \cref{eq:psodeF}, we see that $\RDM$ satisfies the equation $\Upsilon-\omega +\Upsilon V(\RDM)=0$. Thus we make the following definition:

\begin{definition} \label{def:rdm}
Given any spherically symmetric static state we define $\RDM$ to be the radius at which the function $F$ switches from oscillatory to exponential behavior. It satisfies the equation
\begin{equation} \label{eq:rdm}
\Upsilon-\omega +\Upsilon V(\RDM)=0.
\end{equation}
We regard $r=\RDM$ as the edge of the dark matter halo because almost all the dark matter is contained in the region $r\leq\RDM$. See \cref{fig:0,fig:1,fig:10,fig:100}.
\end{definition}

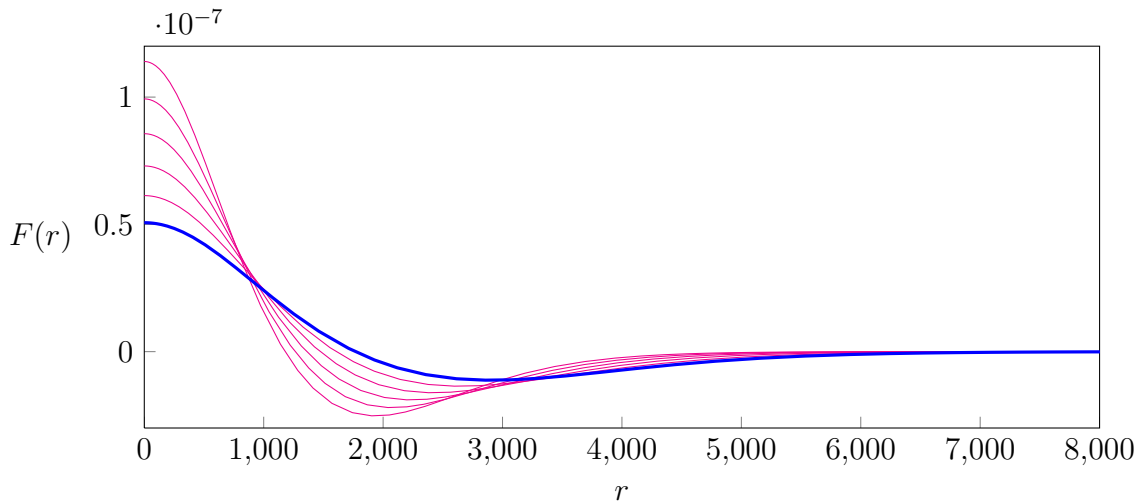
\begin{figure}[hbt]
\centering

\begin{tikzpicture}
\begin{axis}[
	width=5in, height=2in,
	scale only axis,
	xlabel=$r$, ylabel=$F(r)$,
	y label style={rotate=-90},
	xtick pos=left, ytick pos=left,
	xmin=0,xmax=8000,
	ymin=-3e-8, ymax=1.2e-7,
]
\addplot[magenta] table[x expr=\thisrow{r}/1.1,y expr=(1.1)^2*\thisrow{F}]{1.data};
\addplot[magenta] table[x expr=\thisrow{r}/1.2,y expr=(1.2)^2*\thisrow{F}]{1.data};
\addplot[magenta] table[x expr=\thisrow{r}/1.3,y expr=(1.3)^2*\thisrow{F}]{1.data};
\addplot[magenta] table[x expr=\thisrow{r}/1.4,y expr=(1.4)^2*\thisrow{F}]{1.data};
\addplot[magenta] table[x expr=\thisrow{r}/1.5,y expr=(1.5)^2*\thisrow{F}]{1.data};
\addplot[blue,very thick] table[x=r,y=F]{1.data};
\end{axis}
\end{tikzpicture}

\caption{Five scalings of the $n=1$ excited state from \cref{fig:1} (shown in blue) using $\alpha=1.1,1.2,1.3,1.4,1.5$ in the scaling formulas \eqref{eq:psscalings}. Here we only show the dark matter profile function $F(r)$.}
\label{fig:scalings}

\end{figure}

It is well-known that solutions to the PS system \eqref{eq:psode} have certain scaling properties. Specifically, for $\alpha,\beta>0$, we can scale any given solution as follows:
\begin{subequations} \label{eq:psscalings}
\begin{align}
\bar{r} &= \alpha^{-1}\beta^{-1} r \\
\bar{M} &= \alpha\beta^{-3} M \\
\bar{V} &= \alpha^2\beta^{-2} V \\
\bar{F} &= \alpha^2 F \\
\bar{\Upsilon} &= \beta^2\Upsilon \label{eq:mu} \\
(\bar{\Upsilon}-\bar{\omega}) &= \alpha^2(\Upsilon-\omega)
\end{align}
\end{subequations}
and obtain a new solution. These scalings can be derived immediately from \cref{eq:psodeM,eq:psodeV,eq:psodeF}. Note that $\beta$ effectively controls the fundamental constant of nature $\Upsilon$ and $\alpha$ can be used to control the ``peakedness'' of the solution. States which are more compact have higher mass; states which are more spread out have lower mass. In \cref{fig:scalings} we show the effect of scaling a ground state using values of $\alpha$ near $1$. It follows from these scalings that having fixed $\Upsilon$, there is a one-parameter family of static states of order $n$. Once we fix a further characteristic of the solution, a static state is uniquely determined for each $n$, giving us a sequence of static states. We call this ``fixing a scaling'' or ``imposing a scaling condition''.

\section{Tully-Fisher Scalings and Boundary Conditions: Numerical Evidence} \label{sec:numerical}
The basic question we are asking in this paper is:
\begin{question} \label{q:tfscalingsimprecise}
Are there are any scaling conditions for which the sequence of static states $n=0,1,2,\ldots$ obeys a Tully-Fisher-like relation
\begin{equation}
\frac{M}{v^x} = \text{constant}
\end{equation}
for some $3\leq x\leq 4$?
\end{question}
We need to make this question more precise, as it is unclear above what $M$ and $v$ are, exactly, given a particular static state. Astrophysicists attempting to calibrate the baryonic Tully-Fisher relation are faced with a similar problem. In that case, $M$ is an estimate for the total baryonic mass. For $v$, some take it to be the maximum observed velocity; some take it to be an average velocity computed in some well-defined way; some take it to be the velocity observed at a particular (arbitrarily chosen) radius. It is well-known that the rotation curves of many spiral galaxies are somewhat flat, and in this case any of these three choices will produce similar values for $v$.

We have chosen to take $M=M(\RDM)$ and $v=v(\RDM)$. This latter quantity is the circular velocity which would be observed for objects orbiting at $r=\RDM$. This is computed in the standard Newtonian way:
\begin{equation} \label{eq:rotationcurvecalc}
\frac{v(r)^2}{r} = \frac{M(r)}{r^2} \qquad\Longrightarrow\qquad v(r) = \sqrt{\frac{M(r)}{r}}.
\end{equation}
Perhaps a more natural choice for $M$ would be $M_\infty=\lim_{r\to\infty} M(r)$, i.e. the total mass of the static state. But $M(\RDM)\approx M_\infty$ since $\RDM$ is by definition the ``outer edge'' of the halo (see \cref{fig:0,fig:1,fig:10,fig:100}), so it does not really matter which we choose. Similarly, another natural choice for $v$ would be $v_{\text{max}}$, the maximum value of the function $v(r)$. But again, by examining \cref{fig:0,fig:1,fig:10,fig:100}, one can see that $v(\RDM)\approx v_{\text{max}}$. Thus, the following more precise version of \cref{q:tfscalingsimprecise} seems reasonable:
\begin{question} \label{q:tfscalingsprecise}
Are there are any scaling conditions for which the sequence of static states $n=0,1,2,\ldots$ obeys a Tully-Fisher-like relation
\begin{equation} \label{eq:tfscalingsprecise}
\frac{M(\RDM)}{v(\RDM)^x} = \text{constant}
\end{equation}
for some $3\leq x\leq 4$?
\end{question}

Now, in one sense the answer to \cref{q:tfscalingsprecise} is a trivial ``yes'', for we can just pick any $x$ we like and let \cref{eq:tfscalingsprecise} itself be the scaling condition. However, this is obviously cheating; we are interested in scaling conditions which might correspond to something physical. In fact, we have found two conditions that we consider to be more physical. These were given in \cref{sec:intro} along with the Tully-Fisher-like relations thay imply. We restate them here more precisely:
\begin{description}[style=multiline,leftmargin=1.5cm] \label{bcs}
\item[BC1:] Fixing a length scale at $\RDM$ implies a Tully-Fisher-like relation with slope $4$.
\item[BC2:] Fixing $\abs{F(\RDM)}$ implies a Tully-Fisher-like relation with slope $3.4$.
\end{description}

With regards to \textbf{BC1}, we should explain what ``fixing a length scale'' means. We postpone a full explanation to \cref{sec:theoretical} and limit ourselves here to an explanation which involves what we term the ``halflength'' of a state, defined as follows:
\begin{definition} \label{def:halflength}
The $F(r)$ function for any static state has an exponentially decreasing ``tail'' lying to the right of $\RDM$. There is some $R>\RDM$ such that $F(R)=\frac12 F(\RDM)$; we refer to $R-\RDM$ as the \emph{halflength} of the state.
\end{definition}
Requiring the halflength of all states to be the same is one way to fix a length scale at $\RDM$. We refer the reader to \cref{sec:theoretical} for full details.

With regards to \textbf{BC2} note that, by \cref{eq:psodeM}, $\abs{F}$ determines the density of the wave dark matter. This explains why in \cref{sec:intro} we referred to ``fixing the dark matter density'' in \textbf{BC2}.

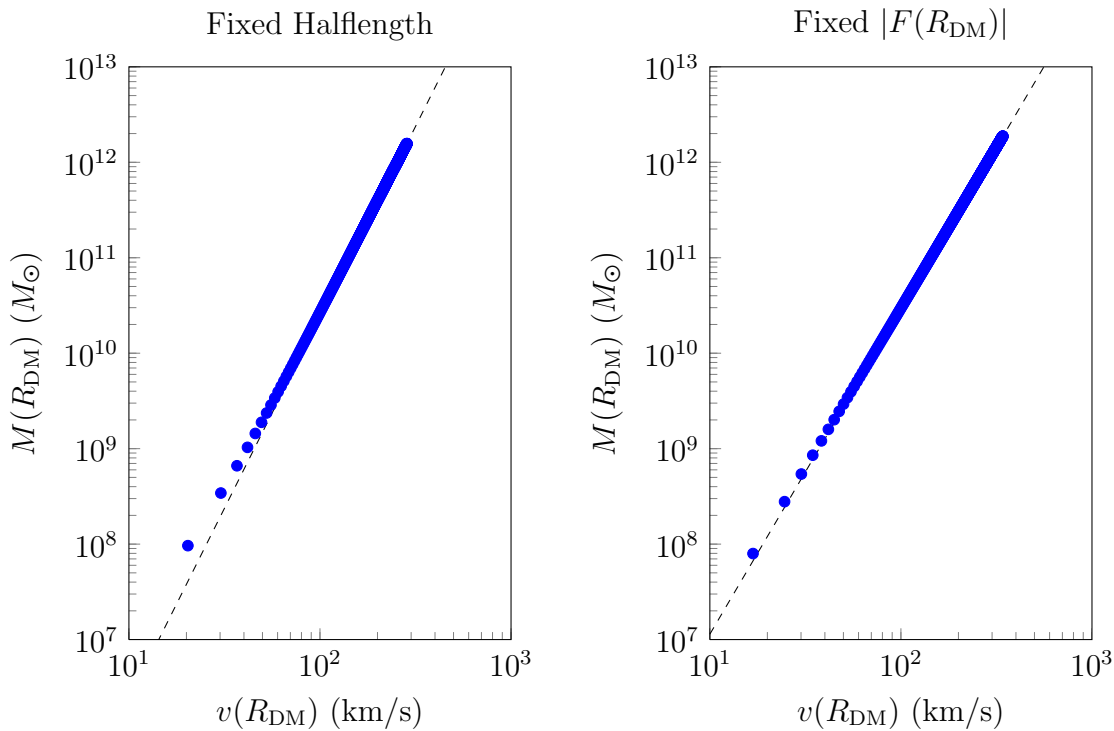
\begin{figure}[hbt]
\centering

\begin{tikzpicture}[baseline]
\begin{loglogaxis}[
	title=Fixed Halflength,
	width=2in, height=3in,
	scale only axis,
	xlabel=$v(\RDM)$ (km/s), ylabel=$M(\RDM)$ ($M_\Sun$),
	xtick pos=left, ytick pos=left,
	xmin=10,xmax=1000,
	ymin=1e7, ymax=1e13,
]
\addplot[only marks,blue] table[x=v,y=M]{fix_halflength.data};
\addplot[black,dashed,domain=10:1000] {3e11*(x/3e5)^4/1.5607e-13};
\end{loglogaxis}
\end{tikzpicture}%
\hspace{0.5cm}%
\begin{tikzpicture}[baseline]
\begin{loglogaxis}[
	title=Fixed $\abs{F(\RDM)}$,
	width=2in, height=3in,
	scale only axis,
	xlabel=$v(\RDM)$ (km/s), ylabel=$M(\RDM)$ ($M_\Sun$),
	xtick pos=left, ytick pos=left,
	xmin=10,xmax=1000,
	ymin=1e7, ymax=1e13,
]
\addplot[only marks,blue] table[x=v,y=M]{fix_F_RDM.data};
\addplot[black,dashed,domain=10:1000] {1.9e22*(x/3e5)^3.4};
\end{loglogaxis}
\end{tikzpicture}

\caption{Numerical evidence for \textbf{BC1} and \textbf{BC2} (see \cpageref{bcs}). At left, the static states $n=0,1,\ldots,800$ with $\Upsilon=\SI{10}{\ly^{-1}}$ and a fixed exponential decay halflength of $\SI{884}{\ly}$. The dashed line is the has slope $4$. At right, the static states $n=0,1,\ldots,800$ with $\Upsilon=\SI{10}{\ly^{-1}}$ and a fixed value for $\abs{F(\RDM)}$ of $\num{2.473e-9}$. The dashed line has slope $3.4$.}
\label{fig:tfscalings}

\end{figure}

It remains in this section to show numerically that \textbf{BC1} and \textbf{BC2} really do give Tully-Fisher-like relations with the stated slopes. We used a computer to derive particular static state solutions for $n=0,1,\ldots,800$ and then scaled these solutions in accordance with \cref{eq:psscalings} so that, first, they all had the same halflength, and second, so they all had the same value for $\abs{F(\RDM)}$. (The precise values for the halflength and $\abs{F(\RDM)}$ were chosen so that the resulting static states would be of reasonable sizes to be representing galactic halos.) The results are shown in \cref{fig:tfscalings}, in which we have plotted $M(\RDM)$ vs. $v(\RDM)$ in log-log space. For large $n$, the static states form a line with the stated slope. (For small $n$, the states deviate slightly from the correct slope. This will be explained in \cref{sec:theoretical}.)

\section{Tully-Fisher Scalings and Boundary Conditions: Theoretical Derivations} \label{sec:theoretical}

In this section we wish to explain theoretically why the statements made in \textbf{BC1} and \textbf{BC2} are true. We have already exhibited their truth numerically in \cref{fig:tfscalings}. 

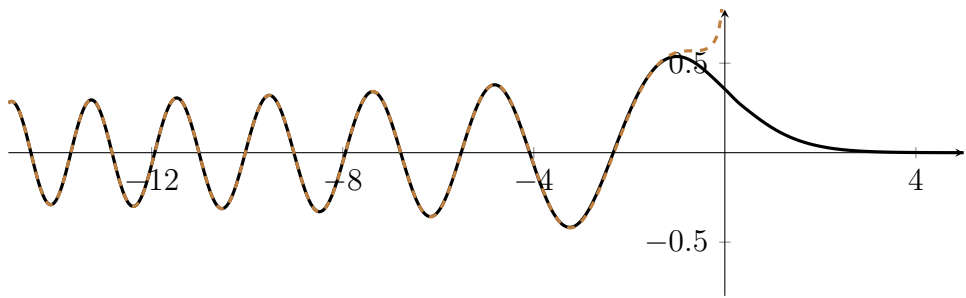
\begin{figure}[hbt]
\centering
\begin{tikzpicture}
\begin{axis}[
	width=5in, height=1.5in,
	scale only axis,
	xmin=-15, xmax=5,
	ymin=-0.8, ymax=0.8,
	xtick={-12,-8,-4,0,4},
	axis lines=middle,
]
\addplot[color=black,solid,very thick,samples=500] gnuplot[id=airy,domain=-15:5] {airy(x)};
\addplot[color=brown,dashed,very thick,samples=500] gnuplot[id=airyapprox,domain=-15:-.000001] {sin(.666666666*(-x)**1.5+(pi/4))/(-x)**.25/sqrt(pi)};
%\addplot[color=gray,dashed,very thick,samples=500] gnuplot[id=airyapprox,domain=-15:-1] {sin(.666666666*(-x)**1.5)};
\end{axis}
\end{tikzpicture}
\caption{In black, the standard solution $\Ai(x)$ to the Airy differential equation $y''=xy$. In brown and dashed, the approximation \eqref{eq:airyapprox} to $\Ai(x)$ for $x<0$.}
\label{fig:airy}
\end{figure}

Both statements involve boundary conditions which fix some aspect of the solutions to the PS system \eqref{eq:psode} around the point $r=\RDM$. We can approximate the solutions around $\RDM$ by using a linear approximation for $V(r)$:
\begin{equation} \label{eq:Vapprox}
\begin{split}
V(r) &\approx V(\RDM) + V'(\RDM)(r-\RDM) \\
&= V(\RDM) + \frac{M(\RDM)}{\RDM^2}(r-\RDM)
\end{split}
\end{equation}
for $r\approx\RDM$. Substituting this approximation for $V$ into \cref{eq:psodeF} and using \cref{eq:rdm}, we find that for $r\approx\RDM$, $F(r)$ approximately solves the differential equation
\begin{equation*}
F_{rr} + \frac{2}{r}F_r = \frac{2\Upsilon^2M(\RDM)}{\RDM^2}(r-\RDM)F.
\end{equation*}
Multiplying through by $r$, we can write this differential equation as
\begin{equation} \label{eq:airyforrF}
(rF)_{rr} = \frac{2\Upsilon^2M(\RDM)}{\RDM^2}(r-\RDM)(rF).
\end{equation}
Compare this to the Airy differential equation $y''=xy$ whose general solution is a linear combination of the two standard functions $\Ai(x)$ and $\Bi(x)$. Both $\Ai(x)$ and $\Bi(x)$ oscillate for negative $x$ and have exponential behavior for positive $x$. The function $\Ai(x)$ is shown in \cref{fig:airy}. It exponentially decreases to zero for $x>0$, whereas the function $\Bi(x)$ (not shown) exponentially increases for $x>0$.

Looking at \cref{eq:airyforrF}, we see that for $r\approx\RDM$, $rF$ approximately solves an equation of the form $y''=a(x-x_0)y$. The general solution to this differential equation is $A\cdot\Ai(a^{1/3}(x-x_0))+B\cdot\Bi(a^{1/3}(x-x_0))$. Since $F$ exponentially decreases for $r>\RDM$, we must have, therefore,
\begin{equation} \label{eq:Fapproxprop}
F(r) \mathrel{\propto} \frac{1}{r} \Ai\left[\left(\frac{2\Upsilon^2M(\RDM)}{\RDM^2}\right)^{1/3}(r-\RDM)\right]
\end{equation}
for $r\approx\RDM$. If we choose the proportionality constant in \cref{eq:Fapproxprop} so that the approximation has the correct value at $r=\RDM$, then we get
\begin{equation} \label{eq:Fapprox}
F(r) \approx F(\RDM)\frac{\RDM}{r} \frac{1}{\Ai(0)} \Ai\left[\left(\frac{2\Upsilon^2M(\RDM)}{\RDM^2}\right)^{1/3}(r-\RDM)\right]
\end{equation}
for $r\approx\RDM$.

\Cref{eq:Fapproxprop} demonstrates mathematically the truth of \textbf{BC1}. For up to a vertical scaling, the shape of $F(r)$ around $r=\RDM$ is given by the expression on the right side of \eqref{eq:Fapproxprop}. Thus, fixing a length scale around $r=\RDM$ evidently amounts to fixing the value of the constant
\begin{equation}
\frac{2\Upsilon^2M(\RDM)}{\RDM^2}.
\end{equation}
This argument explains why fixing a length scale is basically equivalent to fixing a value for the halflength (as we did in \cref{sec:numerical}) or to any of a number of other conditions which fix some length around $\RDM$ (for another example, see our previous paper \cite{bray14}). All such conditions are fixing a particular horizontal scaling of the expression in \cref{eq:Fapprox}.

Now, the rotation curve calculation \eqref{eq:rotationcurvecalc} shows that
\begin{equation} \label{eq:tfconstantatRDM}
\frac{\RDM^2}{M(\RDM)} = \frac{M(\RDM)}{v(\RDM)^4};
\end{equation}
since $\Upsilon$ is a constant (in this paper, $\Upsilon=\SI{10}{\ly^{-1}}$), we see that fixing a length scale is equivalent to fixing the constant in \cref{eq:tfconstantatRDM}. This explains why we get the Tully-Fisher-like relation shown in \cref{fig:tfscalings} with slope $4$. The approximation \eqref{eq:Fapproxprop} gets better for larger $n$ (see \cref{fig:0,fig:1,fig:10,fig:100}), which explains why in \cref{fig:tfscalings} the lower order static states deviate from the slope $4$ line.

To show the truth of \textbf{BC2}, we begin with the approximation \eqref{eq:Fapprox}. This expression contains the four numbers $\Upsilon$, $\RDM$, $M(\RDM)$, and $F(\RDM)$. These cannot all be independently chosen, for the static states effectively live in a 3-parameter space. (Recall the 2-parameter family of scalings \eqref{eq:psscalings}. A third parameter hiding there is the discrete parameter $n$, the order of the static state.) Thus, we should be able to find some relation between these four numbers. In fact, we will argue that
\begin{equation} \label{eq:Fscaling}
\abs{F(\RDM)} \approx 2^{-17/12}\Ai(0)\cdot\Upsilon^{1/6}M(\RDM)^{7/12}\RDM^{-17/12} \qquad\text{for large $n$}
\end{equation}
where
\begin{equation}
2^{-17/12}\Ai(0) \approx \num{0.133}
\end{equation}
This explains \textbf{BC2}, for then we see that fixing $\abs{F(\RDM)}$ amounts to fixing $M(\RDM)^7\RDM^{-17}$. By \cref{eq:rotationcurvecalc}, $\RDM = M(\RDM)v(\RDM)^{-2}$, so fixing $M(\RDM)^7\RDM^{-17}$ is equivalent to fixing $M(\RDM)^{-10}v(\RDM)^{34}=(M(\RDM)/v(\RDM)^{3.4})^{-10}$, i.e. equivalent to fixing
\begin{equation*}
\frac{M(\RDM)}{v(\RDM)^{3.4}}.
\end{equation*}

Our goal now is to derive \cref{eq:Fscaling} beginning with the approximation \eqref{eq:Fapprox}. We will utilize a well-known approximation for the Airy function $\Ai(x)$ for negative inputs:
\begin{equation} \label{eq:airyapprox}
\Ai(-x) \approx \frac{\sin\left(\frac23 x^{3/2}+\frac{\pi}{4}\right)}{\sqrt{\pi}x^{1/4}} \qquad\text{for $x>0$.}
\end{equation}
This approximation can be seen in \cref{fig:airy}. It is extremely good for $x>1$. We will also use the following definite integral, which is easily obtained:
\begin{equation} \label{eq:defint}
\int_0^L \frac{\sin^2\left(\frac23x^{3/2}+\frac{\pi}{4}\right)}{x^{1/2}}\,dx = L^{1/2} + O(1).
\end{equation}
The notation $O(1)$ is mathematical shorthand for a bounded function of $L$.

Using \cref{eq:psodeM,eq:Fapprox}, we have
\begin{equation*}
\begin{split}
M(\RDM) &= \int_0^{\RDM} 8\pi r^2F^2\,dr \\
&\approx \int_0^{\RDM} 8\pi\frac{F(\RDM)^2\RDM^2}{\Ai(0)^2}\Ai\left[-\left(\frac{2\Upsilon^2 M(\RDM)}{\RDM^2}\right)^{1/3}(\RDM-r)\right]^2\,dr \\
&= \int_0^{\RDM} 8\pi\frac{F(\RDM)^2\RDM^2}{\Ai(0)^2}\Ai\left[-\left(\frac{2\Upsilon^2 M(\RDM)}{\RDM^2}\right)^{1/3}r\right]^2\,dr. \\
\end{split}
\end{equation*}
We will temporarily abbreviate $\RDM=R$, $M(\RDM)=M$, $F(\RDM)=F$, $\Ai(0)=A$. Making the subsitution $x=\left(\frac{2\Upsilon^2 M}{R^2}\right)^{1/3}r$ and using the approximation \eqref{eq:airyapprox}, we obtain
\begin{equation*}
\begin{split}
M &\approx 8\pi\frac{F^2R^2}{A^2}\cdot\left(\frac{2\Upsilon^2 M}{R^2}\right)^{-1/3} \int_0^{(2\Upsilon^2MR)^{1/3}} \Ai(-x)^2\,dx \\
&\approx \frac{2^{8/3}}{A^2}\Upsilon^{-2/3}F^2R^{8/3}M^{-1/3} \int_0^{[2\Upsilon^2MR]^{1/3}} \frac{\sin^2\left(\frac23 x^{3/2}+\frac{\pi}{4}\right)}{x^{1/2}}\,dx. \\
\end{split}
\end{equation*}
Using the definite integral \eqref{eq:defint}, we obtain
\begin{equation*}
\begin{split}
M &\approx \frac{2^{8/3}}{A^2}\Upsilon^{-2/3}F^2R^{8/3}M^{-1/3}\left[(2\Upsilon^2MR)^{1/6} + O(1)\right] \\
&\approx \frac{2^{17/6}}{A^2} \Upsilon^{-1/3}F^2R^{17/6}M^{-1/6}.
\end{split}
\end{equation*}
(We dropped the $O(1)$ term because the other term dominates for large $n$.) Rearranging, we get
\begin{equation*}
F^2 \approx 2^{-17/6}A^2\Upsilon^{1/3}M^{7/6}R^{-17/6}.
\end{equation*}
Taking square roots gives us \cref{eq:Fscaling}, which completes the argument.

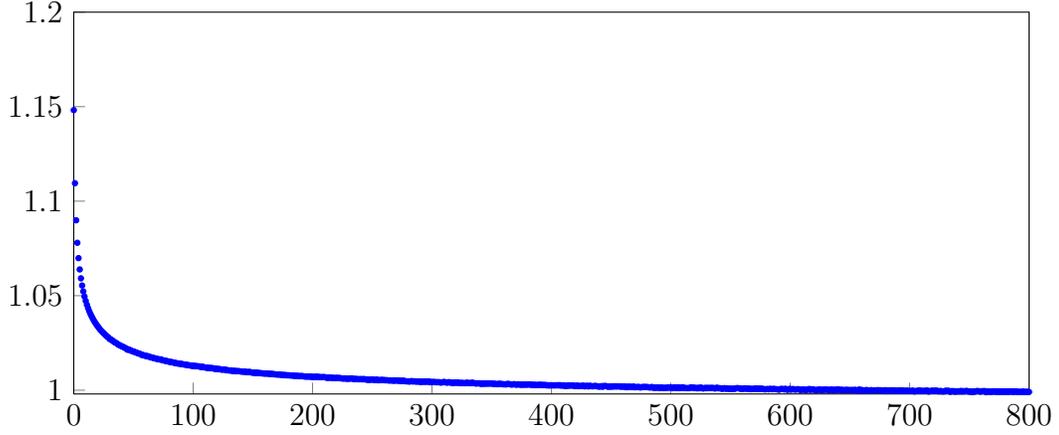
\begin{figure}[p]
\centering

\begin{tikzpicture}
\begin{axis}[
	width=5in, height=2in,
	scale only axis,
	xmin=0, xmax=800,
	ymin=0.998, ymax=1.2,
	xtick pos=left, ytick pos=left,
	xlabel
]
\addplot[only marks,blue,mark size=1pt] table[x=n,y=ratio]{Fscaling.data};
\end{axis}
\end{tikzpicture}

\caption{The ratio given in \cref{eq:Fscalingratio} plotted as a function of $n$ ($n=0,1,\ldots,800$). This graph shows that the approximation \eqref{eq:Fscaling} is very good for large $n$ and not too bad for small $n$.}
\label{fig:Fscaling}
\end{figure}

One might wonder how good the approximation \eqref{eq:Fscaling} is for small $n$. In \cref{fig:Fscaling} we have graphed the ratio
\begin{equation} \label{eq:Fscalingratio}
\frac{\abs{F(\RDM)}}{2^{-17/12}\Ai(0)\cdot\Upsilon^{1/6}M(\RDM)^{7/12}\RDM^{-17/12}}
\end{equation}
as a function of $n$. For the ground state ($n=0$), the ratio is approximately $\num{1.15}$, which is reasonably close to $1$, and for large $n$ the ratio is extremely close to $1$.

Combining \cref{eq:Fapprox,eq:Fscaling} gives the following approximation for the $F(r)$ function of a static state of order $n$. (By convention $F(0)$ is positive so the factor $(-1)^n$ is necessary.)
\begin{equation} \label{eq:Fapprox2}
F(r) \approx (-1)^n2^{-17/12}\Upsilon^{2/12}M(\RDM)^{7/12}\RDM^{-5/12}\cdot\frac{1}{r}\Ai\left[\left(\frac{2\Upsilon^2M(\RDM)}{\RDM^2}\right)^{1/3}(r-\RDM)\right]
\end{equation}
This approximation is shown in \cref{fig:0,fig:1,fig:10,fig:100}.

\section{Other Scaling Conditions} \label{sec:other_scalings}

\begin{figure}[p]
\centering

\begin{tikzpicture}[baseline]
\begin{loglogaxis}[
	title=Fixed $\abs{F(0)}$,
	width=2in, height=3in,
	scale only axis,
	xlabel=$v(\RDM)$ (km/s), ylabel=$M(\RDM)$ ($M_\Sun$),
	xtick pos=left, ytick pos=left,
	xmin=10,xmax=1000,
	ymin=1e7, ymax=1e13,
]
\addplot[only marks,blue] table[x=v,y=M]{fix_F0.data};
\addplot[black,dashed,domain=10:1000] {3e11*(x/3e5)^4/1.5607e-13};
\end{loglogaxis}
\end{tikzpicture}%
\hspace{0.5cm}%
\begin{tikzpicture}[baseline]
\begin{loglogaxis}[
	title=Fixed Location of First Node,
	width=2in, height=3in,
	scale only axis,
	xlabel=$v(\RDM)$ (km/s), ylabel=$M(\RDM)$ ($M_\Sun$),
	xtick pos=left, ytick pos=left,
	xmin=10,xmax=1000,
	ymin=1e7, ymax=1e13,
]
\addplot[only marks,blue] table[x=v,y=M]{fix_first_zero.data};
\addplot[black,dashed,domain=10:1000] {3e11*(x/3e5)^4/1.5607e-13};
\end{loglogaxis}
\end{tikzpicture}

\caption{At left, the static states $n=1,\ldots,800$ with $\Upsilon=\SI{10}{\ly^{-1}}$ and the first node fixed at $\SI{1500}{\ly}$. At right, the static states $n=0,1,\ldots,800$ with $\Upsilon=\SI{10}{\ly^{-1}}$ and $\abs{F(0)}=\num{8e-7}$. The dashed lines have slope $4$. Neither scaling condition gives a Tully-Fisher-like relation.}
\label{fig:other_scalings}

\end{figure}
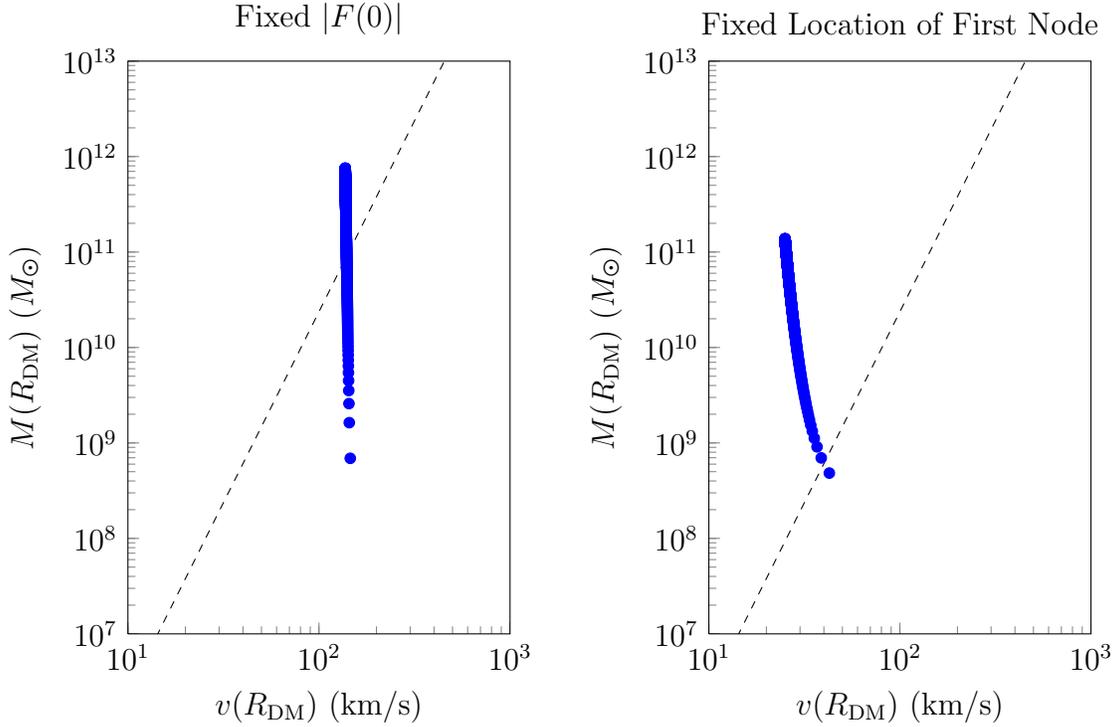

The scaling conditions given in \textbf{BC1} and \textbf{BC2} are boundary conditions imposed at the edge of dark halos. The first imposes a horizontal scale on $F(r)$ and the second imposes a vertical scale. One might wonder whether there are any other scaling conditions which lead to Tully-Fisher-like relations. For example, what happens when we try the same thing at the center of the halo? To impose a horizontal scale on $F(r)$ at the origin, we can fix the location of the first node (for $n\geq 1$). To impose a vertical scaling, we can fix $\abs{F(0)}$. In \cref{fig:other_scalings} we show the results of imposing these two scaling conditions. Neither leads to a Tully-Fisher-like relation. In fact, besides \textbf{BC1} and \textbf{BC2} no other scaling condition we have tried leads to a Tully-Fisher-like relation.

\section{Conclusion} \label{sec:conclusion}
We have exhibited two boundary conditions, \textbf{BC1} and \textbf{BC2}, which we restate here once more in their more colloquial form from \cref{sec:intro}:
\begin{description}[style=multiline,leftmargin=1.5cm]
\item[BC1:] Fixing a length scale at the outer edge of spherically symmetric wave dark matter halos implies a Tully-Fisher-like relation with slope $4$.
\item[BC2:] Fixing the dark matter density at the outer edge of spherically symmetric wave dark matter halos implies a Tully-Fisher-like relation with slope $3.4$.
\end{description}
We gave both numerical evidence and theoretical arguments illustrating the truth of \textbf{BC1} and \textbf{BC2}. Our main goal in this paper has been to demonstrate that a Tully-Fisher-like relation can arise naturally out of the equations of wave dark matter.

To conclude we comment briefly on the connection between a Tully-Fisher-like relation for wave dark matter halos and the observed baryonic Tully-Fisher relation. Define the \emph{baryon fraction} of a galaxy to be the fraction of the galaxy's mass which is baryonic. If the baryon fraction were the same for every galaxy, then the baryonic Tully-Fisher relation would have the same slope as a Tully-Fisher-like relation for dark halos. However, it is observed that the baryon fraction of galaxies decreases with size. Small galaxies contain mostly dark matter, and only very large galaxies contain baryons in sufficient quantities to approach the cosmic baryon fraction, which is around $1/6$. This is known as the \emph{missing baryon problem} (see \cite{mcgaugh10}).

In $M$ vs. $v$ log-log space, when switching from plotting dark mass to baryonic mass, all galaxies shift vertically downward. The missing baryon problem implies that large galaxies shift less than small galaxies. Therefore, if the baryonic Tully-Fisher relation is related to a Tully-Fisher-like relation for dark halos, the former relation should have a steeper slope than the latter.

\pgfplotsset{
	axisoptions/.style={width=5in,
		height=1.5in,
		scale only axis,
		y label style={rotate=-90},
		ytick pos=left,
		extra x tick labels={},
		extra x tick style={grid=major,tickwidth=0pt},
	},
	plotoptions/.style={thick},
	axisoptionsM/.style={ylabel=$M(r)$,
		xtick={\empty},
		scaled x ticks=false,
		scaled y ticks=true,
		},
	axisoptionsV/.style={ylabel=$V(r)$,
		xtick={\empty},
		scaled x ticks=false,
		scaled y ticks=true,
		},
	axisoptionsF/.style={ylabel=$F(r)$,
		xtick={\empty},
		scaled x ticks=false,
		scaled y ticks=true,
	},
	axisoptionsv/.style={ylabel=$v(r)$,
		xlabel={$r$},
		xtick pos=left,
		scaled y ticks=true,
		},
	plotoptionsM/.style={color=red},
	plotoptionsV/.style={color=olive},
	plotoptionsF/.style={color=blue},
	plotoptionsv/.style={color=orange}
}
\pgfplotsset{
	axisoptions0/.style={
		xmin=0,xmax=8800,
		extra x ticks=2452,
		height=1.3in,
	},
}

\begin{figure}[p]
\centering

\begin{tikzpicture}
% M(r)
\begin{axis}[
	name=M,
	axisoptions,
	axisoptionsM,
	axisoptions0,
	ymin=-1e-8, ymax=2.6e-5,
]
\addplot[plotoptions,plotoptionsM] table[x=r,y=M]{0.data};
\end{axis}

% V(r)
\begin{axis}[
	name=V,
	at=(M.below south west),
	anchor=above north west,
	axisoptions,
	axisoptionsV,
	axisoptions0,
	ymin=-2e-8, ymax=1e-11,
]
\addplot[plotoptions,plotoptionsV] table[x=r,y=V]{0.data};
\addplot[plotoptions,plotoptionsV,dashed] table[x=r,y=Va]{0.data};
\end{axis}

% F(r)
\begin{axis}[
	name=F,
	at=(V.below south west),
	anchor=above north west,
	axisoptions,
	axisoptionsF,
	axisoptions0,
	ymin=-1e-10, ymax=3e-8,
]
\addplot[plotoptions,plotoptionsF] table[x=r,y=F]{0.data};
\addplot[plotoptions,plotoptionsF,dashed] table[x=r,y=Fa]{0.data};
\end{axis}

% v(r) (rotation curve)
\begin{axis}[
	name=v,
	at=(F.below south west),
	anchor=above north west,
	axisoptions,
	axisoptionsv,
	axisoptions0,
	ymin=-1e-8, ymax=1e-4,
]
\addplot[plotoptions,plotoptionsv] table[x=r,y=v]{0.data};
\end{axis}
\end{tikzpicture}

\caption{A ground state ($n=0$) with $\Upsilon=\SI{10}{\ly^{-1}}$. Units are geometrized and given in light-years. The first three functions graphed are the solutions $M(r)$, $V(r)$, $F(r)$ to \cref{eq:psode}. They are, respectively, the total mass profile, the Newtonian potential, and the dark matter profile. The fourth function is a rotation curve $v(r)$ calculated in accordance with \cref{eq:rotationcurvecalc}. The vertical gray line marks the location of $\RDM$. The dashed curves show the approximations described in \cref{sec:theoretical} in \cref{eq:Vapprox,eq:Fapprox2}.}
\label{fig:0}

\end{figure}
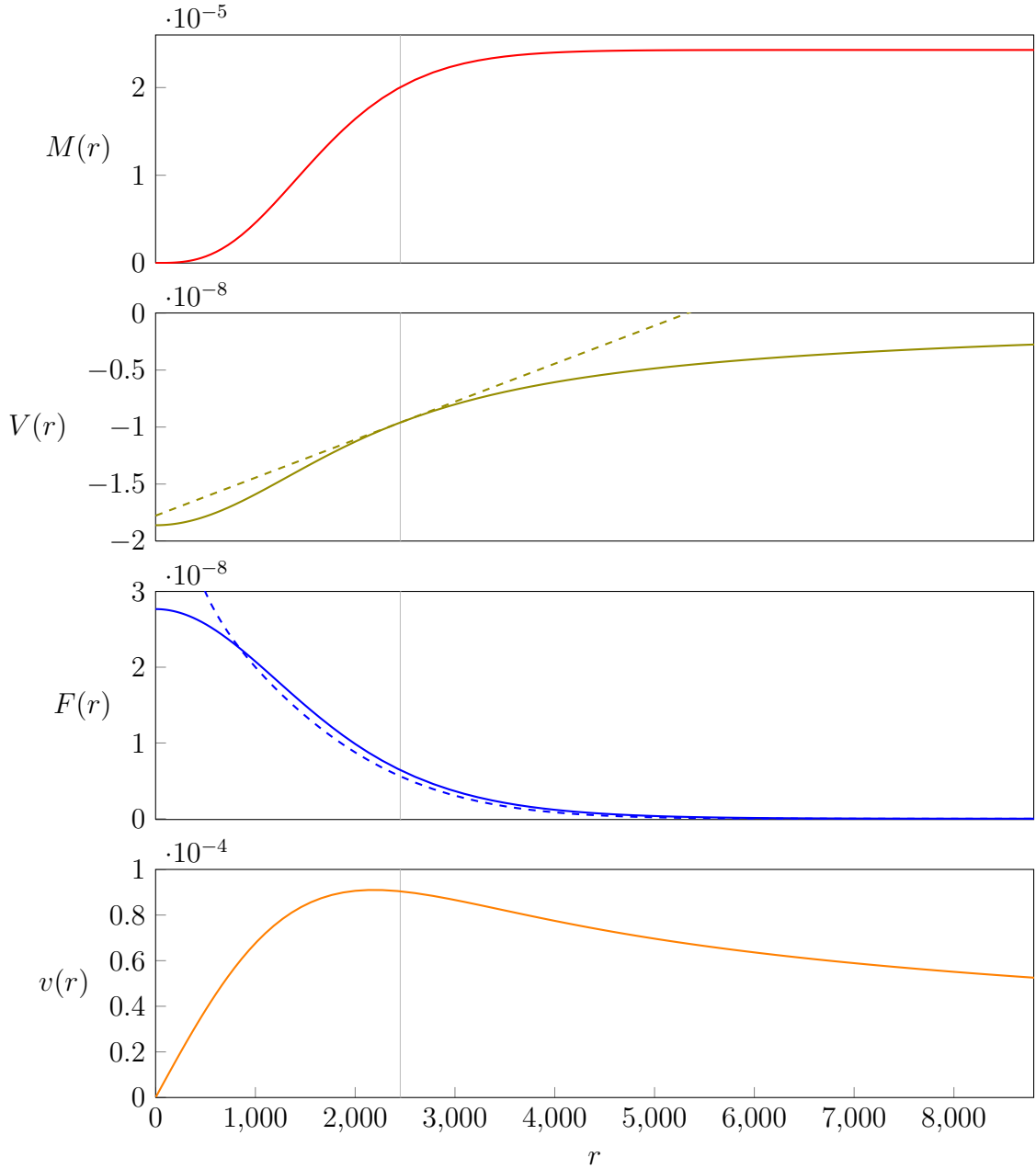

\pgfplotsset{
	axisoptions0/.style={
		xmin=0,xmax=12000,
		extra x ticks=4384,
	},
}

\begin{figure}[p]
\centering

% M(r)
\begin{tikzpicture}
\begin{axis}[
	name=M,
	axisoptions,
	axisoptionsM,
	axisoptions0,
	ymin=-1e-9, ymax=7.7e-5,
]
\addplot[plotoptions,plotoptionsM] table[x=r,y=M]{1.data};
\end{axis}

% V(r)
\begin{axis}[
	name=V,
	at=(M.below south west),
	anchor=above north west,
	axisoptions,
	axisoptionsV,
	axisoptions0,
	ymin=-4.5e-8, ymax=1e-12,
]
\addplot[plotoptions,plotoptionsV] table[x=r,y=V]{1.data};
\addplot[plotoptions,plotoptionsV,dashed] table[x=r,y=Va]{1.data};
\end{axis}

% F(r)
\begin{axis}[
	name=F,
	at=(V.below south west),
	anchor=above north west,
	axisoptions,
	axisoptionsF,
	axisoptions0,
	ymin=-2e-8, ymax=6e-8,
]
\addplot[plotoptions,plotoptionsF] table[x=r,y=F]{1.data};
\addplot[plotoptions,plotoptionsF,dashed] table[x=r,y=Fa]{1.data};
\end{axis}

% v(r) (rotation curve)
\begin{axis}[
	name=v,
	at=(F.below south west),
	anchor=above north west,
	axisoptions,
	axisoptionsv,
	axisoptions0,
	ymin=-1e-8, ymax=1.5e-4,
]
\addplot[plotoptions,plotoptionsv] table[x=r,y=v]{1.data};
\end{axis}
\end{tikzpicture}

\caption{An excited state ($n=1$). See the caption to \cref{fig:0} for a fuller description.}
\label{fig:1}

\end{figure}
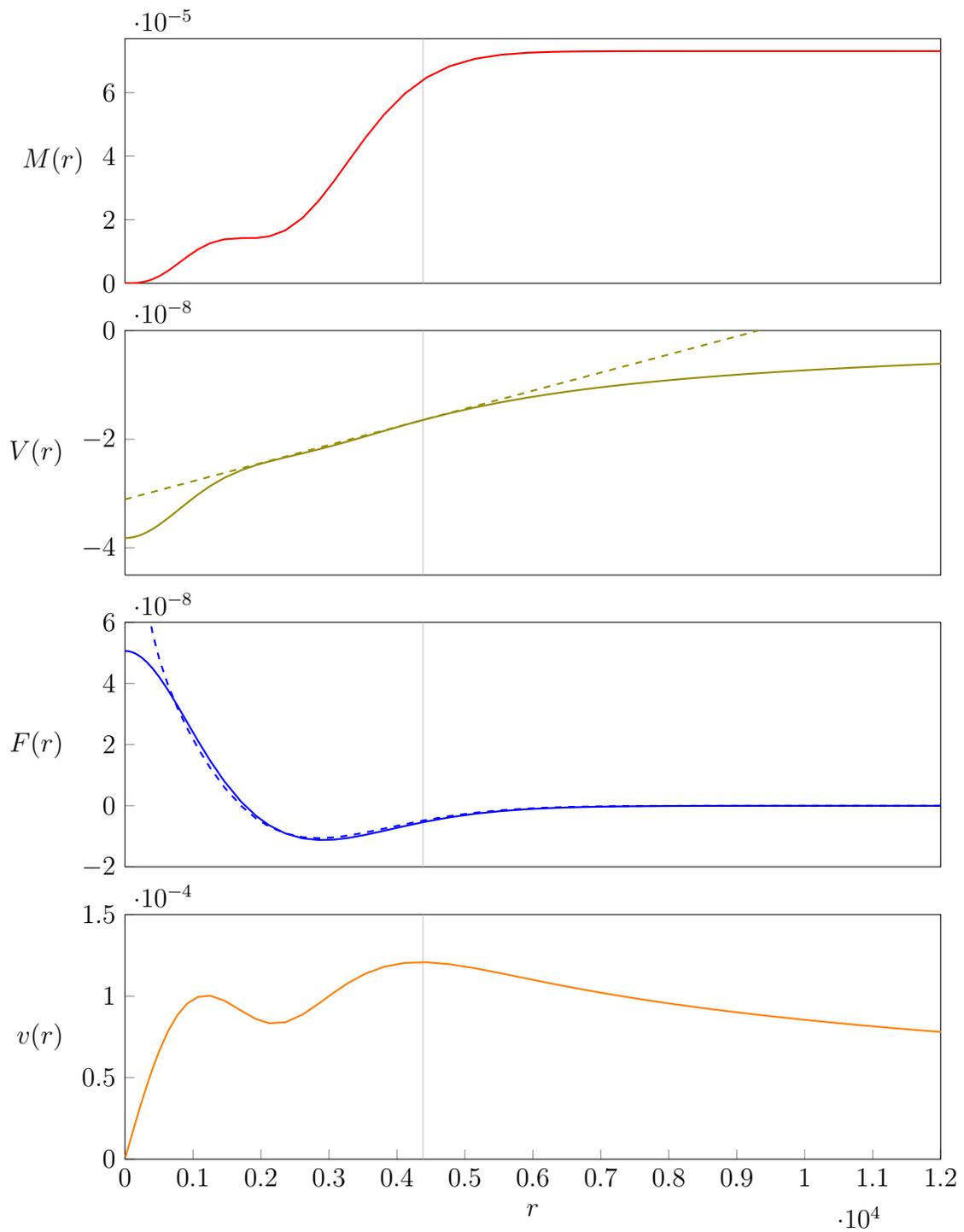
\pgfplotsset{
	axisoptions0/.style={
		xmin=0,xmax=22000,
		extra x ticks=15061,
	},
}

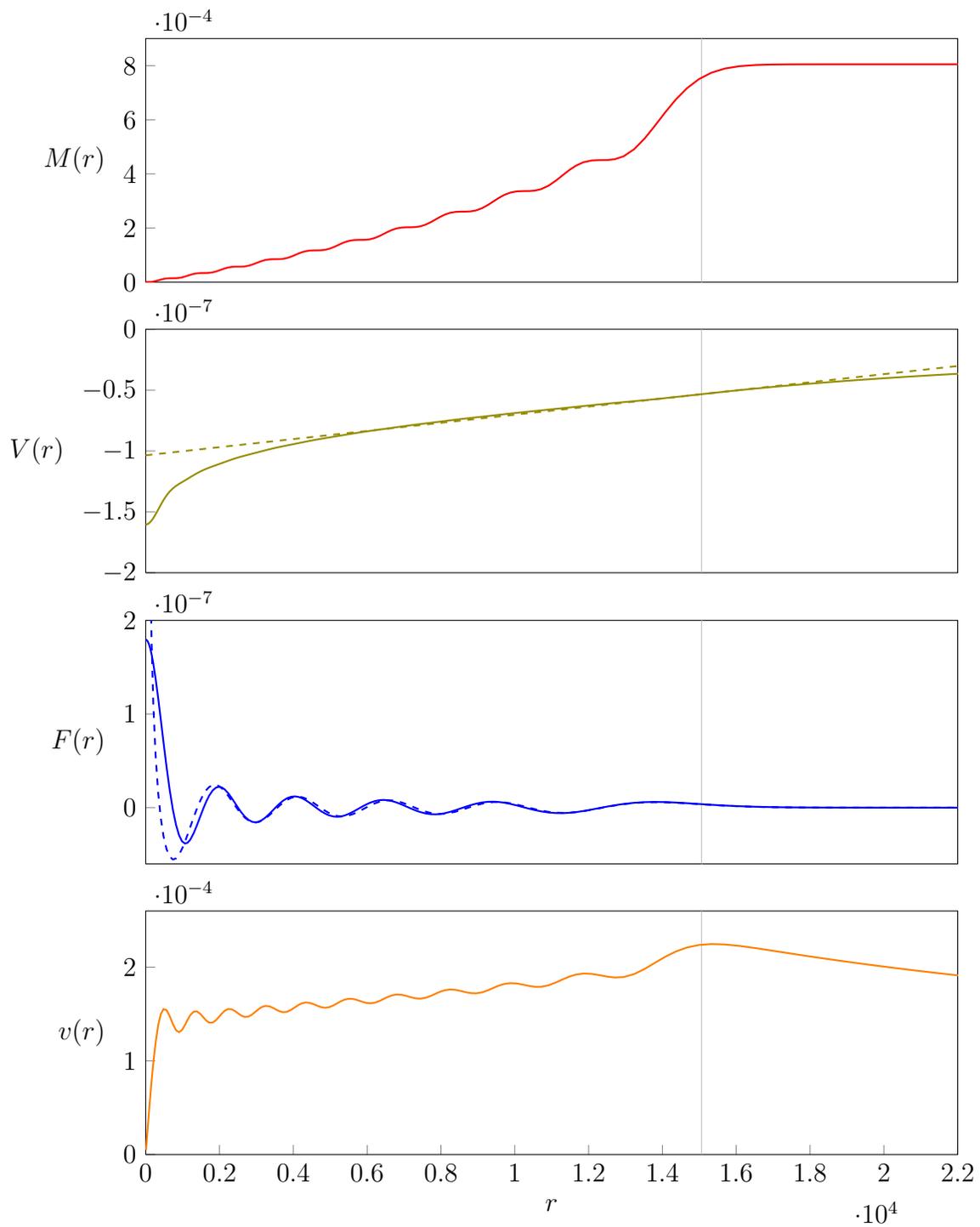
\begin{figure}[p]
\centering

% M(r)
\begin{tikzpicture}
\begin{axis}[
	name=M,
	axisoptions,
	axisoptionsM,
	axisoptions0,
	ymin=-1e-9, ymax=9e-4,
]
\addplot[plotoptions,plotoptionsM] table[x=r,y=M]{10.data};
\end{axis}

% V(r)
\begin{axis}[
	name=V,
	at=(M.below south west),
	anchor=above north west,
	axisoptions,
	axisoptionsV,
	axisoptions0,
	ymin=-2e-7, ymax=1e-12,
]
\addplot[plotoptions,plotoptionsV] table[x=r,y=V]{10.data};
\addplot[plotoptions,plotoptionsV,dashed] table[x=r,y=Va]{10.data};
\end{axis}

% F(r)
\begin{axis}[
	name=F,
	at=(V.below south west),
	anchor=above north west,
	axisoptions,
	axisoptionsF,
	axisoptions0,
	ymin=-6e-8, ymax=2e-7,
]
\addplot[plotoptions,plotoptionsF] table[x=r,y=F]{10.data};
\addplot[plotoptions,plotoptionsF,dashed] table[x=r,y=Fa]{10.data};
\end{axis}

% v(r) (rotation curve)
\begin{axis}[
	name=v,
	at=(F.below south west),
	anchor=above north west,
	axisoptions,
	axisoptionsv,
	axisoptions0,
	ymin=-1e-8, ymax=2.6e-4,
]
\addplot[plotoptions,plotoptionsv] table[x=r,y=v]{10.data};
\end{axis}
\end{tikzpicture}

\caption{An excited state ($n=10$). See the caption to \cref{fig:0} for a fuller description.}
\label{fig:10}

\end{figure}
\pgfplotsset{
	axisoptions0/.style={
		xmin=0,xmax=73000,
		extra x ticks=67742,
	},
}

\begin{figure}[p]
\centering

\begin{tikzpicture}
% M(r)
\begin{axis}[
	name=M,
	axisoptions,
	axisoptionsM,
	axisoptions0,
	ymin=-1e-6, ymax=1.8e-2,
]
\addplot[plotoptions,plotoptionsM] table[x=r,y=M]{100.data};
\end{axis}

% V(r)
\begin{axis}[
	name=V,
	at=(M.below south west),
	anchor=above north west,
	axisoptions,
	axisoptionsV,
	axisoptions0,
	ymin=-1e-6, ymax=1e-11,
]
\addplot[plotoptions,plotoptionsV] table[x=r,y=V]{100.data};
\addplot[plotoptions,plotoptionsV,dashed] table[x=r,y=Va]{100.data};
\end{axis}

% F(r)
\begin{axis}[
	name=F,
	at=(V.below south west),
	anchor=above north west,
	axisoptions,
	axisoptionsF,
	axisoptions0,
	ymin=-5e-8, ymax=5e-8,
]
\addplot[plotoptions,plotoptionsF] table[x=r,y=F]{100.data};
\addplot[plotoptions,plotoptionsF,dashed] table[x=r,y=Fa]{100.data};
\end{axis}

% v(r) (rotation curve)
\begin{axis}[
	name=v,
	at=(F.below south west),
	anchor=above north west,
	axisoptions,
	axisoptionsv,
	axisoptions0,
	ymin=-1e-9, ymax=5.5e-4,
]
\addplot[plotoptions,plotoptionsv] table[x=r,y=v]{100.data};
\end{axis}

\end{tikzpicture}

\caption{An excited state ($n=100$). See the caption to \cref{fig:0} for a fuller description. The top and bottom of the $F(r)$ function have been cropped out to show more detail.}
\label{fig:100}

\end{figure}
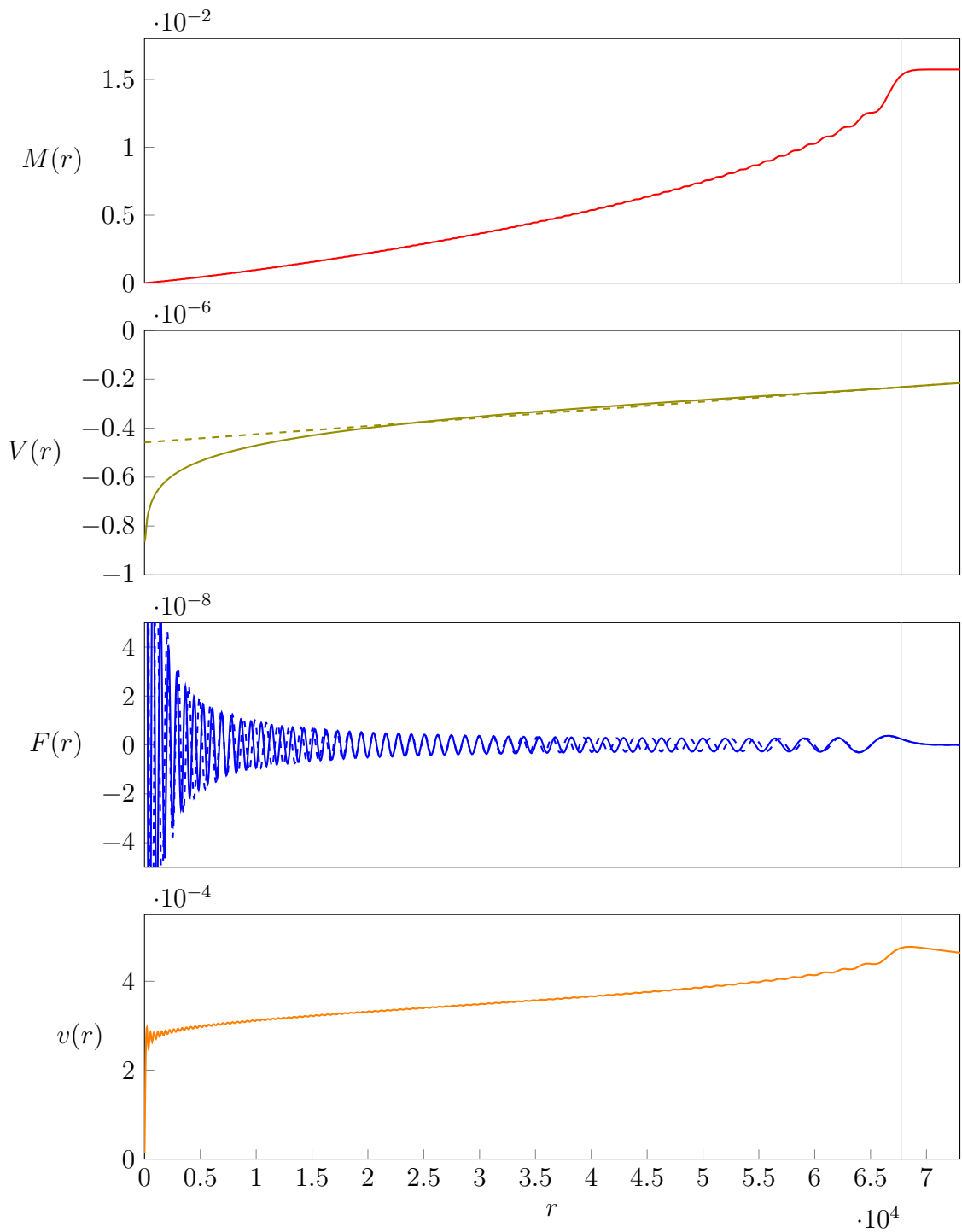

\newpage
\printbibliography

\end{document}